\documentclass[%
 reprint,
 superscriptaddress,
 amsmath,amssymb,
 aps,
]{revtex4-2}

\usepackage{graphicx}
\usepackage{hyperref}
\usepackage{xfrac}
\usepackage{placeins}
\usepackage{siunitx}
\usepackage{mathtools}
\usepackage{amsmath} 

\hypersetup{colorlinks, urlcolor=blue, linkcolor=blue}

\DeclarePairedDelimiter\ket{\lvert}{\rangle}
\newcommand{\microsecond}{\si{\micro\second}}

\begin{document}

\title{Optical Detection of Coherent Nuclear Spin Control in a Molecular System at Room Temperature}

\author{Eoin O. Irving}
\affiliation{James Watt School of Engineering, University of Glasgow, Glasgow, G12 8QQ, UK.}

\author{Ziqiu Huang}
\affiliation{Department of Materials, Imperial College London, Prince Consort Road, London, SW7 2AZ, UK}

\author{Max Attwood}
\email{m.attwood@imperial.ac.uk}
\affiliation{Department of Materials, Imperial College London, Prince Consort Road, London, SW7 2AZ, UK}

\author{Sam L. Bayliss}
\email{sam.bayliss@glasgow.ac.uk}
\affiliation{James Watt School of Engineering, University of Glasgow, Glasgow, G12 8QQ, UK.}

\author{Sarah K. Mann}
\email{sarah.mann@glasgow.ac.uk}
\affiliation{James Watt School of Engineering, University of Glasgow, Glasgow, G12 8QQ, UK.}

\begin{abstract}
Molecular systems with optically addressable electron spins are emerging as versatile quantum sensing platforms, offering tunable qubit properties and molecular-scale spatial control. Owing to their longer lifetimes, coupled nuclear spins provide an attractive resource for enhancing sensing performance. For molecular spin systems, the ability to initialise, coherently control and readout nuclear spins at room temperature is crucial for realising such benefits. Here, we present the first optical detection of nuclear magnetic resonance and coherent nuclear spin control with a molecular system at room temperature. Specifically, we demonstrate optically detected Rabi oscillations of \textsuperscript{14}N nuclear spins in the photoexcited triplet state of the organic chromophore 6,13-diazapentacene. The strong hyperfine coupling between the electronic spin and \textsuperscript{14}N nuclei enhances the nuclear Rabi frequency, enabling fast coherent nuclear-spin rotations. This work lays the foundation for leveraging nuclear spins as memories for enhanced molecular-based quantum sensing. 
\end{abstract}
\maketitle

Nuclear spins are a key resource in quantum science owing to their long lifetimes, making them attractive as memories for preserving quantum information~\cite{Maurer1second2012, Wang10hours2025}. In contrast, electron spins possess magnetic moments that are orders of magnitude larger, enabling stronger coupling to magnetic fields and other physical quantities in sensing applications. Furthermore, electronic spin states can be initialised and read out optically, facilitating single-spin detection and room-temperature operation~\cite{Gruber1997ScanningCenters, Suter2017Single-spinDiamond}. As a result, optically addressable electronic spins have emerged as an important platform for quantum technologies, particularly quantum sensing~\cite{schirhagl2014nitrogen}. Coupled electron--nuclear spin systems offer a route to combine the advantages of both.     

Opportunities with electron--nuclear spin systems have been explored extensively with solid state defects, particularly with the nitrogen-vacancy (NV$^{-}$) centre in diamond. For example, coupling between the NV$^-$ electron spin and a nearby nuclear spin allows repetitive electron spin readout, enabled by the much longer nuclear spin lifetime ~\cite{jiang2009repetitive}. This technique improves the sensitivity of optically addressable spin sensors~\cite{arunkumar2023quantum} and allowed single shot readout of a nuclear spin state~\cite{neumann2010single}. Sensitivity enhancing techniques of electron--nuclear registers, combined with the high spatial resolution of NV$^{-}$ sensors, has enabled nanoscale nuclear magnetic resonance at ambient temperatures~\cite{Lovchinsky2016NuclearLogic, Aslam2017NanoscaleResolution}. Beyond the NV$^-$, room temperature manipulation of optically readable electron--nuclear systems has also been demonstrated in hexagonal boron nitride spin defects~\cite{Gao2022NuclearNitride, Tibiassy2026CoherentNitride}.


Complementing solid-state spin defects, optically addressable \textit{molecular} spins combine quantum functionality with the tunability of synthetic chemistry. A molecular construction enables precise control over optical and electronic properties~\cite{bayliss_optically_2020, laorenza_tunable_2021, bayliss_enhancing_2022}, while offering versatile routes to integration and deployment~\cite{mann2026optically}. A wide range of molecular systems that exhibit optically addressable spin states have been demonstrated, including coordination complexes that host ground state-spins~\cite{bayliss_optically_2020}, organic materials such as diradicals~\cite{Kopp2024LuminescentQubits, chowdhury_bright_2025} and carbenes~\cite{RoggorsCarbenes2025}, and photoexcited spins in fluorescent proteins~\cite{federfluorescentprotein2025, abrahamsquantum2026} and organic chromophores~\cite{tesimanSurveyingOpticallyAddressable2026b}. Among these, pentacene has played a pioneering role, from the first demonstrations of optically detected single spins~\cite{wrachtrup_optical_1993, kohler_magnetic_1993}, to, more recently, the first demonstrations of room-temperature optically detected coherent control of molecular spins~\cite{Mena2024Room-TemperatureSpins, Singh2025Room-temperatureCrystals}, pressure and temperature coupling exceeding the NV$^-$ performance~\cite{singhHighSensitivityPressure2025}, vector AC field sensing~\cite{li2025robustacvectorsensing}, and integration with 2D material surfaces~\cite{Zhou2Dsurfaces2026, zheng2026surfacescaffoldedmolecularqubit}. 

\begin{figure*}[t!]
    \includegraphics{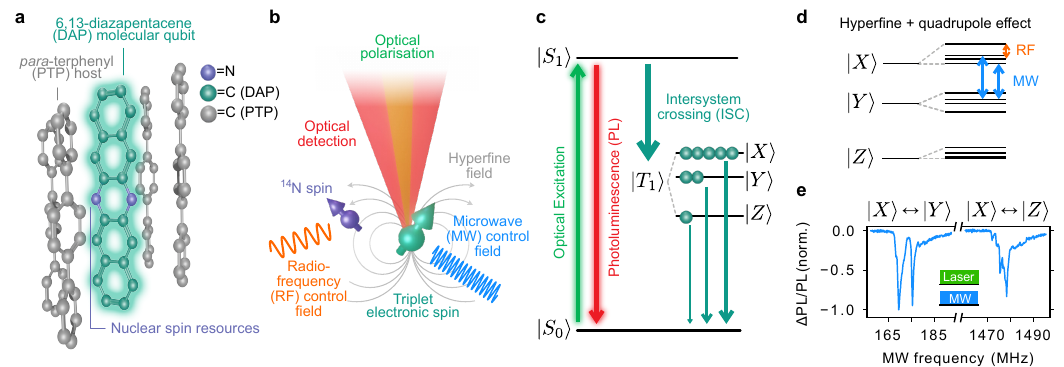} 
    \centering
    \caption{\textbf{A room-temperature optically addressable electron--nuclear molecular system.} \textbf{a} Chemical structure of DAP in a PTP crystal host. \textbf{b} Schematic illustration of the coupled electron--nuclear spin system. The electron spin is optically polarised and read out, while radio frequency (RF) and microwave (MW) pulses drive NMR and electron spin resonance, respectively. \textbf{c} Energy level diagram of DAP. Optical excitation of the $\ket{S_0}\rightarrow\ket{S_1}$ transition is followed by intersystem crossing into the triplet manifold, anisotropically populating the $\ket{T_1}$ sublevels. Differential relaxation of these sublevels to $\ket{S_0}$ enables optical spin readout. \textbf{d} Hyperfine and quadrupolar coupling of the two $I=1$ \textsuperscript{14}N nuclei splits each triplet sublevel. \textbf{e} Zero-field continuous-wave ODMR spectrum of the $\ket{X} \leftrightarrow \ket{Y}$ and  $\ket{X} \leftrightarrow \ket{Z}$ transitions in a DAP:PTP single crystal, with splittings due to the \textsuperscript{14}N couplings (data reproduced from Ref.~\cite{Mann2025ChemicallyResonance}).}
  \vspace{2mm} 
  \hrule
  \vspace{1mm}
  \label{Fig1}
\end{figure*}

Yet to be explored with optically addressable molecular spins is
access to nuclei at room temperature. Such room-temperature electron--nuclear spin registers would expand molecular functionality for quantum sensing, enabling enhanced sensing protocols as exemplified by solid-state defects. 
Furthermore, a molecular synthesis approach opens up deterministic control of nuclear spin placement, atomic species, and hyperfine coupling. Requirements for fabrication of a molecular room temperature spin register are (i) optical detection of nuclear magnetic resonance (NMR) and (ii) optically detected coherent control of the nuclear spin states, both yet to be realised. Optically detected NMR (OD-NMR) of nuclei coupled to molecular spins has been demonstrated at cryogenic temperatures~\cite{chanOpticalDetectionEndor1969, HarrisODENDOR1969, Dinse1978DeterminationODNQR}, including at the sample size limit of a single nuclear spin~\cite{wrachtrup1997magnetic}. More recently, optically detected coherent nuclear spin control has been demonstrated in a europium-based molecular complex (without an electron spin), also under cryogenic conditions~\cite{Vasilenko2026OpticallyComplex}.
Here we demonstrate the first OD-NMR and optically detected coherent nuclear spin control with a molecular system at room temperature. 

To realise this, we use a \textsuperscript{14}N-substituted pentacene derivative, 6,13-Diazapentacene (DAP), doped in a \textit{para}-terphenyl (PTP) crystal (Figure \ref{Fig1}a). Previously, we demonstrated that the \textsuperscript{14}N substitution of pentacene to form DAP improved the optical-spin contrast from $\sim$~20~\% to 40~\%~\cite{Mann2025ChemicallyResonance}. In this work, we manipulate the $I=1$ \textsuperscript{14}N nuclear spins present in DAP. The \textsuperscript{14}N spin states are optically addressable via the hyperfine coupling to the electron spin (Figure \ref{Fig1}b). We use these couplings to perform optically detected electron nuclear double resonance (OD-ENDOR), measuring the \textsuperscript{14}N NMR transitions and demonstrating optically detected nuclear Rabi oscillations at room temperature. The strong hyperfine interaction enables enhanced nuclear Rabi frequencies, facilitating fast nuclear-spin manipulations. These results enhance the potential of molecular spins for quantum sensing applications, by combining room-temperature operation, optical addressability, and coherent nuclear-spin control within a chemically tunable platform.

\section*{Results}
\subsection*{Optically Detected Nuclear Magnetic Resonance}

All measurements were performed at room temperature. 
Photoexcitation of DAP creates a spin-polarised triplet state through spin-selective intersystem crossing from the first excited singlet state ($\ket{S_1}$; Figure~\ref{Fig1}c). In the absence of an external magnetic field, the resulting triplet sublevels---denoted $\ket{X}$, $\ket{Y}$ and $\ket{Z}$---are split by the zero-field splitting, defined by parameters $D=1390.4~\mathrm{MHz}$ and $E=-84.9~\mathrm{MHz}$~\cite{Mann2025ChemicallyResonance}. Here, the $X$, $Y$, and $Z$ axes of DAP correspond to the molecule's long, short and out-of-plane directions, respectively. Spin-selectivity of the intersystem crossing results in initial relative triplet populations of $P_X:P_Y:P_Z \simeq 0.743:0.143:0.114$ \cite{Mann2025ChemicallyResonance}. Microwave (MW) excitation resonant with transitions between the triplet sublevels redistributes the spin populations, leading to a change in the photoluminescence (PL) intensity through spin-dependent relaxation and re-excitation pathways, enabling optically detected magnetic resonance (ODMR). 

Hyperfine and quadrupolar coupling of the \textsuperscript{14}N nuclei further splits the electron sublevels (Figure \ref{Fig1}d). The resulting eigenfunctions at zero-field are superpositions of product states, $\ket{iv_1v_2}$, where $i \in \{X, Y, Z\}$; and $v_1, v_2 \in \{x, y, z\}$ are eigenstates of the \textsuperscript{14}N quadrupolar Hamiltonian. The state mixing is mediated by the hyperfine interaction
\begin{equation}
\hat{H}_{\mathrm{hf}}=\sum_{j=1}^{2}\left(A_{xx}\hat{S}_x\hat{I}_{jx}+A_{yy}\hat{S}_y\hat{I}_{jy}+A_{zz}\hat{S}_z\hat{I}_{jz}\right)
\end{equation}
where $\{A_{kk}\}$ are the diagonal hyperfine tensor elements and $\hat{S}_k$, $\hat{I}_{jk}$ are the spin operator elements for the electronic and two \textsuperscript{14}N spins, respectively ($k\in\{x, y, z\}$). See the supplementary text and Table S1 for a full description of the interactions and resulting zero-field eigenstates.
Experimentally, these interactions are evident in the $\ket{X}\leftrightarrow\ket{Y}$ and $\ket{X}\leftrightarrow\ket{Z}$ ODMR spectra shown in Figure~\ref{Fig1}e, where the strong nuclear couplings give rise to multiple spectral features. This resolved splitting enables nuclear-spin transitions to be selectively addressed, forming the basis of the OD-NMR measurements presented below.

\begin{figure}[!h]
    \includegraphics{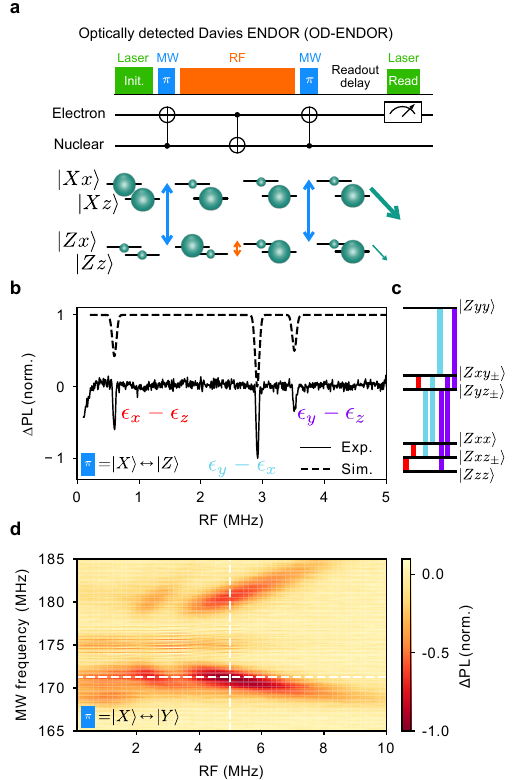} 
    \centering
    \caption{\textbf{Room temperature optically detected nuclear magnetic resonance}. 
    \textbf{a} OD-ENDOR pulse sequence. The equivalent quantum gates and relative populations at each stage of the sequence are also depicted.  
    \textbf{b} \textsuperscript{14}N OD-ENDOR spectrum of DAP:PTP for a selective MW $\pi$-pulse on the $\ket{X}\leftrightarrow\ket{Z}$ transition. Peak assignments correspond to the transitions indicated on the energy-level diagram in \textbf{c} (see Table S1 for eigenstate compositions). 
    \textbf{d} OD-ENDOR map measured by sweeping both the MW frequency around the $\ket{X}\leftrightarrow\ket{Y}$ transition and the RF in the OD-ENDOR sequence. The white dashed crosshair marks the frequency pairing used to drive nuclear Rabi oscillations (see Figure \ref{Rabis}).} 
    \label{ENDOR}
\end{figure}

For OD-NMR of the \textsuperscript{14}N spins we use optically detected Davies ENDOR~\cite{Davies1974ATechnique}, depicted in Figure~\ref{ENDOR}a. The pulse sequence is accompanied by the quantum gate sequence, together with a schematic representation of the spin-state populations throughout the experiment, illustrated by the green spheres. Following optical initialisation, a selective MW $\pi$-pulse creates nuclear spin polarisation within an electron sublevel (in quantum logic terminology, this is a CNOT gate with a nuclear control and electronic target). 
An RF pulse then drives transitions between the nuclear spin states (a CNOT gate where now the electron is the control and the nuclear spin is the target). A second MW $\pi$-pulse tries to reverse the effect of the first MW pulse, mapping any changes induced by the RF pulse onto a difference in electron spin populations. A short delay repopulates the ground state via spin-dependent relaxation and this is read out optically using a second laser pulse. OD-ENDOR contrast is determined by comparing the signal with and without an RF pulse. 

We apply a selective MW pulse on the $\ket{X}\leftrightarrow\ket{Z}$ transition and sweep the RF to produce the OD-ENDOR spectrum shown in Figure~\ref{ENDOR}b. Three resonances are observed at 0.60, 2.91, and 3.51 MHz. In this experiment, the RF duration (20~\microsecond) exceeds the lifetime of the short-lived $|X \rangle$ manifold ($1/k_X = 4$~\microsecond~\cite{Mann2025ChemicallyResonance}), so any initial $\ket{X}$ population decays before the OD-ENDOR sequence is completed. Consequently, the measured spectrum originates from splittings of the longer-lived $\ket{Z}$ sublevel. The OD-ENDOR signal exhibits negative contrast, corresponding to a decrease in PL intensity upon the application of a resonant RF pulse, as this results in less population being recovered to the brighter $\ket{X}$ state.

The resulting OD-ENDOR spectrum corresponds primarily to the \textsuperscript{14}N quadrupolar structure within the $\ket{Z}$ electronic manifold. The first-order hyperfine interaction vanishes since $\langle i|\hat{H}_{\mathrm{hf}}|i \rangle=0$ for $\ket{i} \in \ket{X}, \ket{Y}, \ket{Z}$ leaving the second-order interactions as the leading terms. From previous density functional calculations, we take the hyperfine terms to be 
$A_{xx}, A_{yy}, A_{zz} = -0.793, -0.988, 23.2$~MHz~\cite{Mann2025ChemicallyResonance}. 
For $\ket{Z}$ manifold states the second-order interactions are $\sim A_{xx}^2/D$ and $A_{yy}^2/D$, thus $\sim1$ kHz - weak compared to the quadrupole interaction of $\sim4$ MHz (see below). We therefore assign the transitions at 0.61, 2.92, and 3.51 MHz to the quadrupolar eigenstate transitions $\epsilon_x - \epsilon_z$, $\epsilon_y - \epsilon_x$ and $\epsilon_y - \epsilon_z$, respectively, where $\{\epsilon_i\}$ are the eigenvalues of the \textsuperscript{14}N quadrupolar Hamiltonian.
The assigned NMR transitions are indicated on the energy-level diagram in Figure~\ref{ENDOR}c.

\begin{figure*}[!ht]
    \includegraphics{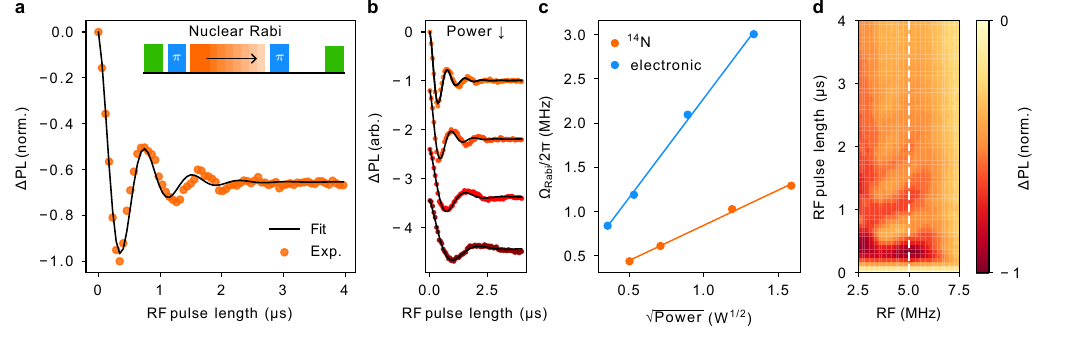} 
    \centering
    \caption{\textbf{Room-temperature optically detected coherent nuclear spin control in a molecular system}. 
    \textbf{a} Coherent nuclear Rabi oscillations fitted to a single decaying sinusoid, with the OD-ENDOR pulse sequence shown in the inset.
    \textbf{b} RF power dependence of nuclear Rabi oscillations. The triplet decay exponential was subtracted from all time traces for plotting and extraction of Rabi frequencies via FFTs. 
    \textbf{c} RF power dependence of the Rabi frequency for the nuclear Rabi oscillations in \textbf{b}, compared to the MW power dependence of electronic Rabi frequencies on the $\ket{X} \leftrightarrow \ket{Y}$ transition. Solid lines are linear fits. \textbf{d} RF detuning dependence of the nuclear Rabi oscillations. The white dashed line is the RF used in \textbf{a} and \textbf{b}.}
    \vspace{2mm} 
    \hrule
    \vspace{1mm}
    \label{Rabis}
\end{figure*}

Fitting the OD-ENDOR spectrum (Figure~\ref{ENDOR}b; dashed black line) yields $\epsilon_{x}, \epsilon_{y}, \epsilon_{z} = -0.77, 2.14, -1.37$~MHz (defined in the molecular frame, see the supplementary text for simulation details). 
These values correspond to a quadrupolar coupling constant, $|e^2qQ/h| = 2\epsilon_y = 4.28$~MHz, and an asymmetry parameter, $\eta = (\epsilon_x - \epsilon_z)/\epsilon_y = 0.28$. These parameters closely match those reported for the structurally related three-benzene-ring system phenazine-\textit{d}\textsubscript{8}, for which $|e^2qQ/h| = 4.36$~MHz and $\eta = 0.21$~\cite{Dinse1978DeterminationODNQR}. The experimentally determined quadrupolar coupling constant is also in close agreement with the density functional theory value $|e^2qQ/h| = 4.44$~MHz, while the calculated asymmetry parameter ($\eta = 0.11$) somewhat underestimates the degree of asymmetry in the electric field gradient~\cite{Mann2025ChemicallyResonance}. To our knowledge, these results are the first room temperature demonstration of OD-NMR in a molecular system.

\subsection*{Optically Detected Coherent Nuclear Spin Control} 

Having demonstrated OD-NMR, we next investigate coherent nuclear-spin control. While the transitions we measured coupled to the $\ket{Z}$ sublevel have drive rates corresponding to $\pi$-times on the order of the triplet sublevel lifetime $1/k_Z=50~\microsecond$, we can use the stronger hyperfine mixing present across the $\ket{X}$ and $\ket{Y}$ manifolds to enhance the nuclear drive rate (which we discuss further below). We first observe the OD-ENDOR behaviour around the $\ket{X}\leftrightarrow\ket{Y}$ transition by producing a two-dimensional OD-ENDOR map of MW and radio frequency, shown in Figure~\ref{ENDOR}d. The spectrum is produced with a $720$~ns RF pulse, contrasting the $20~\microsecond$ RF pulse required for the $\ket{Z}$ manifold OD-ENDOR spectrum (Figure~\ref{ENDOR}b). The broad features observed and comparatively short RF pulse required is indicative of strong coupling to the driving RF field. The map also shows pronounced diagonal structures, consistent with contributions from multiple overlapping proton transitions whose frequencies depend on the selected MW excitation (see supplementary text for further discussion). 

We then select the resonant MW and RF indicated in Figure~\ref{ENDOR}c, and by varying the RF pulse duration, drive coherent nuclear Rabi oscillations within the electronic $\ket{X}$ and $\ket{Y}$ manifolds (Figure~\ref{Rabis}a). This demonstrates optically detected coherent nuclear spin control at room temperature and opens a key capability for molecular systems by establishing optically readable nuclear-spin qubits under ambient conditions.

In Figure~\ref{Rabis}b we vary the RF amplitude to demonstrate the expected linear dependence of the nuclear Rabi frequency, and compare the slope to that measured for electronic Rabi oscillations between the $\ket{X}$ and $\ket{Y}$ states (Figure~\ref{Rabis}c). Rabi frequencies are extracted from the fast Fourier transforms (FFTs) of the data (see Figure~S1 for FFTs of Figure~\ref{Rabis}b, and Figure~S2 for electronic $\ket{X}\leftrightarrow\ket{Y}$ Rabi oscillations and FFTs). 
A linear fit yields slopes of 0.8 and 2.2 MHz W$^{-1/2}$, for the nuclear and electronic Rabi frequencies, respectively.
Figure~\ref{Rabis}c shows the RF detuning dependence of the Rabi oscillations. Detuning down from the resonance (white dashed line) demonstrates the expected increase in oscillation frequency. The asymmetry in the detuning pattern reflects the OD-ENDOR line shape (see Figure~\ref{ENDOR}d).

The Rabi frequencies we observe are on the order of $\sim$ MHz.
Using the experimentally determined electronic Rabi frequency of $\Omega_e = 3.0$~MHz and the ratio of \textsuperscript{14}N to electronic gyromagnetic ratios, the expected direct nuclear drive frequency is estimated to be only $\Omega_{^{14}\mathrm{N}} = \Omega_e \gamma_{^{14}\mathrm{N}}/\gamma_e \approx 330$ Hz. 
The enhancement of approximately three orders of magnitude arises from hyperfine-mediated mixing of electron and nuclear spin states~\cite{Sangtawesin2016Hyperfine-enhancedDiamond}. 

As discussed above, hyperfine mixing is negligible within the $\ket{Z}$ manifold; however, the $\ket{X}$ and $\ket{Y}$ manifolds are mixed through $A_{zz}\hat{S}_z\hat{I}_z$ by the second-order perturbation of $\sim A_{zz}^2/2E$. Consequently, nominally nuclear-spin transitions acquire partial electronic character and couple much more strongly to the applied RF field, producing MHz-scale nuclear Rabi oscillations. Calculations of the RF drive Hamiltonian matrix elements predict an enhancement of approximately three orders of magnitude for transitions within the electronic $\ket{X}$- and $\ket{Y}$-manifolds compared to a hyperfine-free nuclear spin (see the supplementary text). 

This enhancement of Rabi frequency directly corresponds to a decrease in quantum logic gate times and thus the sensor overhead time in electron--nuclear spin sensing schemes. However, the increase in coupling to the applied RF field naturally brings an increase in coupling to other fields contributing to nuclear decoherence~\cite{Sangtawesin2016Hyperfine-enhancedDiamond}. 
Working within a molecular framework provides future opportunities for selective nuclear substitution, thus the ability to tune the hyperfine coupling and find the optimal balance between gate times and decoherence. Similarly, sample deuteration would decrease the influence of the system's magnetic environment, which may be significant given the observed proton contributions to the OD-ENDOR map, potentially enabling higher fidelity nuclear spin control. Additionally, relaxation of the electronic spin to its ground-singlet state $(S=0)$ may provide a pathway to extension of the nuclear spin coherence, by effectively `turning off' the electron spin and its contribution to nuclear decoherence~\cite{brownCoherentStateTransfer2011, filidou2012ultrafast}.

Given we are operating at Rabi rates approaching the resonant nuclear transition, the hyperfine enhancement provides an accessible platform for investigation into the strong driving regime and time-optimised quantum control. This merit has been demonstrated in previous work including sub-nanosecond electron spin flips in an NV$^-$ centre~\cite{fuchsGigahertz2009} and has prompted RF antenna design optimisation for achieving strongly driven \textsuperscript{1}H Rabi oscillations~\cite{yudilevichCoherentManipulationNuclear2023}.

\section*{Discussion}

We present potential future directions from this work for spin-based sensing at room temperature.
The synthetic tunability of molecular systems provides a route to engineering the nuclear spin environment of the optically addressable molecular system. 
As discussed above, control over the hyperfine interaction allows for maintaining the desired balance between enhanced nuclear gyromagnetic ratios and nuclear coherence.
This hyperfine tunability also enables optimisation of the nuclear relaxation, key for sensing protocols that employ nuclear spins as quantum memories. 
Then, systems of multiple distinct optically addressable nuclear spins could increase the quantum information storage capacity of a single molecule while enabling more advanced quantum sensing protocols~\cite{Giovannetti2004Quantum-EnhancedLimit} and quantum error correction~\cite{Waldherr2014QuantumRegister}. 

An additional opportunity arises in the context of photoexcited triplets such as DAP, which possess spin-free electronic ground states. 
A permanently present electron spin can limit nuclear-spin coherence and lifetimes through the hyperfine interaction. 
In photoexcited systems, however, the hyperfine interaction can be switched on and off through optical excitation and relaxation~\cite{filidou2012ultrafast, Lee2013ReadoutAncilla}. 
The electron spin can therefore be activated only when required---for example, to mediate fast gate operations or enable optical readout---while the spin-free ground state preserves long nuclear coherence, providing an attractive platform for nuclear-spin-assisted memory protocols.

The demonstration of OD-NMR here represents an important step toward the optical detection of nuclei beyond the host molecule itself. Access to nuclear spins could enhance the spectral resolution of nanoscale NMR through nuclear-spin correlation spectroscopy, where the achievable resolution is no longer limited by the electron-spin lifetime~\cite{Laraoui2013High-resolutionDiamond}.
Molecular architectures such as thin films~\cite{Mena2024Room-TemperatureSpins}, metal-organic frameworks~\cite{Yamauchi2024ModulationFramework} and molecules scaffolded on 2D material surfaces~\cite{Zhou2Dsurfaces2026, zheng2026surfacescaffoldedmolecularqubit} offer routes to controlled integration of the sensor in close proximity to target spins; and established ENDOR-based spin label-techniques for measuring molecular distances could be extended to optical readout~\cite{Meyer2020MeasurementSpectroscopy, Sielaff2026PulsedScale}. 
In addition, the combination of OD-NMR with pentacene-based platforms, which are highly efficient agents for room-temperature dynamic nuclear polarisation (DNP)~\cite{Henstra1990HighTemperature, Kouno2019}, could open directions for proposed fundamental physics tests using pentacene nanoparticles~\cite{steiner2025optically}.

Overall, our demonstration of optically detected coherent nuclear spin control at room temperature combined with the synthetic tunability of a molecular platform establishes a foundation for engineering tailored electron--nuclear spin systems for quantum science and technology.

\section*{Methods}\label{methods}
\subsection*{Sample Preparation}
DAP was synthesised as previously described~\cite{Ng2023}. DAP:PTP (0.05\%) single crystals were grown using the Bridgman method, following the procedure reported previously in Ref.~\cite{Mann2025ChemicallyResonance}.

\subsection*{OD-NMR}
All experiments described were performed at room-temperature and zero-applied external magnetic field. The sample was excited with a 520 nm laser diode (Thorlabs LP520-SF40) pulsed using a pulse generator (Agilent 8114A) triggered by an external waveform generator (Swabian Instruments Pulse Streamer 8/2). The laser polarization was adjusted with a half-wave plate (Thorlabs AHWP05M-520) to optimize the PL. The laser light was filtered through a  bandpass filter (Thorlabs FBH520-40), directed towards the sample with a 550 nm long-pass dichroic mirror (Thorlabs DMLP550R) and focused onto the sample with an objective lens (Nikon CFI S Plan Fluor ELWD 20XC). The same objective lens collected the resultant PL which passed through the dichroic, separating it from the laser path, before further filtering through a 550 nm long-pass (Thorlabs FELH550). An achromatic doublet (AC254-030-AB) was used to focus the PL onto a multimode fiber coupled to a photodetector (FEMTO OE-200-SI).

Both MW and RF oscillating fields were generated with separate signal generators (Windfreak SYNTHNVPRO for MW, Stanford SG396 for RF) which were each externally gated with a microwave switch (Minicircuits ZASWA-2-50DRA). The switches were toggled by the same Pulse Streamer 8/2. When exciting the $\ket{X}\leftrightarrow\ket{Y}$ electronic transition, the signals were combined with a two-way power splitter/combiner (Minicircuits ZMSC-2-1+) and the combined signal was amplified by a high power amplifier (Minicircuits LZY-22+). For the $\ket{X}\leftrightarrow\ket{Z}$ electronic transition, the MW signal was separately amplified (Minicircuits ZHL-25W-272+) and the signals were combined with a diplexer (Microwave Circuits D5506991). 
In each case the signal was transmitted onto the sample using a single-turn loop of wire. The loop was mounted on a three-axis translation stage to optimise MW and RF delivery to the sample.
The sample was loaded onto a rotation mount to optimise the OD-ENDOR signal by adjusting the alignment of the molecular transition dipoles and the oscillating field direction. 
The voltage output from the photodetector was both recorded and demodulated with a lock-in amplifier (Stanford SR830). The lock-in amplifier measured the difference in PL between measurement and control pulse sequences alternated at the reference frequency (generated by the Pulse Streamer 8/2). For OD-ENDOR, the control sequence was obtained with the RF switched off.

In the OD-ENDOR sequence, for all measurements, we used a 5$~\microsecond$ laser pulse to initialise the triplet state, a  $5~\microsecond$ readout delay to allow spin-selective $\ket{S_0}$ repopulation, a $5~\microsecond$ laser pulse to read out the repopulation and a $50~\microsecond$ delay before the next sequence iteration to allow the system to reset to $\ket{S_0}$. For the $\ket{Z}$ OD-ENDOR spectra, the MW amplitude was tuned to give 350 ns selective $\pi$-pulses, centred at 1.4771 GHz. For the $\ket{X}\leftrightarrow\ket{Y}$ OD-ENDOR map and the nuclear Rabi oscillations we used 340 ns selective $\pi$-pulses centred at 171.3 MHz.

\subsection*{Data availability}
The data underlying this work are available at [URL to be added in proof].

\subsection*{Acknowledgments}
This work was supported by UK Research and Innovation [grant numbers MR/W006928/1, UKRI3889, EP/W027542/1]. S.K.M acknowledges support from a Glasgow Engineering Futures Fellowship. For the purpose of open access, the authors have applied a Creative Commons Attribution (CC BY) licence to any Author Accepted Manuscript version arising from this submission.

\subsection*{Author contributions}
S.L.B and S.K.M initiated the project. E.I and S.K.M performed ODMR measurements and analysed data, with input from S.L.B on interpretation. Z.H and M.A prepared samples. M.A, S.L.B and S.K.M provided supervision and acquired funding. E.I and S.K.M wrote the manuscript with input from all authors.
 
\subsection*{Competing interests}
The authors declare that they have no competing interests.

\subsection*{Additional Information}
\subsubsection*{Supplementary Materials}
Supplementary materials are available at [URL to be added in proof].

\nocite{EasySpin} \nocite{Sloop1981ElectronCrystals}

%

\end{document}